\documentclass[aps,prl,twocolumn,floatfix,showpacs]{revtex4}
\usepackage{amsmath}
\usepackage{graphicx}
\usepackage{amsfonts}
\usepackage{color}

\renewcommand{\aa}{{\bf a}}
\newcommand{\bb}{{\bf b}}
\newcommand{\GG}{{\bf G}}
\newcommand{\xx}{{\bf x}}
\newcommand{\rr}{{\bf r}}
\newcommand{\kk}{{\bf k}}
\newcommand{\gn}{\nabla n}

\begin{document}

\title{Efficient implementation of a van der Waals density functional:
       Application to double-wall carbon nanotubes}

\author{ Guillermo Rom\'an-P\'erez and Jos\'e M. Soler }
\affiliation{ Departamento de F\'{\i}sica de la Materia Condensada, 
              C-III, Universidad Aut\'{o}noma de Madrid,
              E-28049 Madrid, Spain }

\date{\today}

\begin{abstract}
   
   We present an efficient implementation of the van der Waals density
functional of Dion {\it et al} [Phys. Rev. Lett. {\bf 92}, 246401 (2004)], 
which expresses the nonlocal correlation energy as a double spacial 
integral.
   We factorize the integration kernel and use fast Fourier transforms 
to evaluate the selfconsistent potential, total energy, and atomic forces,
in $N \log N$ operations.
   The resulting overhead in total computational cost, over semilocal 
functionals, is very moderate for medium and large systems.
   We apply the method to calculate the binding energies and the barriers
for relative translation and rotation in double-wall carbon nanotubes.

\end{abstract}



\maketitle

   Density functional theory (DFT) has become the method of choice
for first-principles simulations of static and dynamical properties
of complex materials with strong ionic, covalent, and metallic 
interactions.
   However, weak van der Waals (vdW) interactions are also essential for 
many systems and processes, like molecular solids and liquids, surface 
adsorption, and biological reactions \cite{MullerDethlefs-Hobza2000}.
   Local or semilocal density functionals obviously cannot describe 
asymptotically the nonlocal dispersion correlations.
   At binding distances, they have been frequently found to give 
reasonable results~\cite{PerezJorda-Becke1995,Tsuzuki-Luthi2001} but 
their ability to do so is generally very sensitive to the specific
functional used and its parametrization details, what makes the 
``{\it ab initio}'' character of this approach rather questionable.
   Thus, the simulation of vdW systems has typically relied on atom-atom 
potentials with the conventional~\cite{Dobson2006} $r^{-6}$ asymptotic 
behavior and with parameters fitted to empirical data or to accurate 
quantum chemistry calculations of simple molecules.
   Such potentials are also added as plug-ins to {\it ab initio} semilocal 
density functionals \cite{Elstner2001,Grimme2004}.
   Another approach includes vdW interactions through
effective atom-electron pseudopotentials \cite{vonLilienfeld2004}.
   However, the accuracy and reliability of such approaches is
limited because vdW energies arise from electron-electron correlations
that depend not only on the atomic species but also on their chemical
environment.
   More {\it ab initio} wavefunction-dependent approaches are more
reliable but also much more expensive \cite{Angyan2005}.

   Thus, a key development has been the proposal by 
Dion {\it et al}~\cite{Dion2004} of a universal nonlocal energy 
functional of the electron density $n(\rr)$ with the form
\begin{equation}
E_{xc}[n(\rr)] = E_x^{GGA}[n(\rr)] + E_c^{LDA}[n(\rr)] + E_c^{nl}[n(\rr)]
\label{Exc}
\end{equation}
where the exchange energy $E_x^{GGA}$ is described through the semilocal
generalized gradient approximation (GGA)~\cite{Zhang-Yang1998} and the
correlation energy has a local part $E_c^{LDA}$, described in the local 
density approximation (LDA), and a nonlocal (nl) part $E_c^{nl}$ given by
\begin{equation}
E_c^{nl}[n(\rr)] = \frac{1}{2} \int\int d^3\rr_1 ~d^3\rr_2 ~n(\rr_1) ~n(\rr_2) 
   ~\phi(q_1,q_2,r_{12})
\label{Enl}
\end{equation}
where $r_{12}=|\rr_1-\rr_2|$, and $q_1, q_2$ are the values of a 
universal function $q_0(n(\rr),|\gn(\rr)|)$, evaluated at $\rr_1$ and $\rr_2$.
   The kernel $\phi$ has also a precise and universal 
form that in fact depends only on two variables $d_1=q_1 r_{12}$ and 
$d_2=q_2 r_{12}$, but it can obviously be written also as a function of 
$q_1, q_2$, and $r_{12}$, what we will find convenient.
   The shape of $\phi$ obeys that: 
{\it i}) $E_c^{nl}$ is strictly zero for any system with constant density; 
and {\it ii}) the interaction between any two molecules has the correct 
$r^{-6}$ dependence for large separations $r$.
   Using a direct evaluation of Eq.(\ref{Enl}), 
this vdW functional has been applied successfully to a variety of
systems, including interactions between pairs of atoms and molecules,
molecules adsorbed on surfaces, molecular solids, and biological systems.
\cite{Dion2004,Thonhauser2006,ChakarovaKack2006,Kleis2007,Cooper2008}

   If $q_1$ and $q_2$ in Eq.~(\ref{Enl}) were fixed values, independent 
of $\rr_1$ and $\rr_2$, $E_c^{nl}$ would be a simple convolution, like
the Coulomb energy, that could be evaluated by Fourier methods.
   Therefore, our key step for an efficient implementation is to expand 
the kernel $\phi$ as
\begin{equation}
\phi(q_1,q_2,r_{12}) \simeq \sum_{\alpha \beta} 
   \phi(q_{\alpha},q_{\beta},r_{12}) ~p_{\alpha}(q_1) ~p_{\beta}(q_2)
\label{phi}
\end{equation}
where $q_{\alpha}$ are {\em fixed} values, chosen to ensure a good 
interpolation of function $\phi$.
In order to illustrate how the factorization~(\ref{phi}) can be performed in a
systematic way, we consider first the interpolation of a function $f(x)$
using a linear scheme, like those of Lagrange, Fourier, or splines:
\begin{equation}
f(x) \simeq \sum_{\alpha} f_\alpha ~p_{\alpha}(x)
\label{fofx}
\end{equation}
where $f_\alpha=f(x_\alpha)$ and $p_{\alpha}(x)$ is the function 
resulting from the interpolation of the particular values 
$f_{\beta} = \delta_{\alpha \beta}$.
   In Lagrange interpolation, it is a polynomial of given order.
   In Fourier interpolation it has the form 
$~\sin(\pi(x-x_{\alpha})/\Delta x)/(\pi(x-x_{\alpha})/\Delta x)$.
   We use cubic splines, in which $p_{\alpha}(x)$ is a succession of 
cubic polynomials in every interval $[x_{\beta},x_{\beta+1}]$, 
matching in value and first two derivatives at every point $x_{\beta}$.
   Notice that $p_{\alpha}(x)$ depends on the interpolation scheme and
on the (fixed) points $x_{\alpha}$, but not on the interpolated function.
   In two-dimensional interpolation, one typically interpolates first
in one variable and then in the other:
\begin{eqnarray}
f(x,y) &\simeq& \sum_{\beta} f(x,y_\beta) ~p_{\beta}(y) \nonumber \\
       &\simeq& \sum_\beta
  \left( \sum_{\alpha} f(x_\alpha,y_{\beta}) 
          ~p_{\alpha}(x) \right) p_\beta(y)
\label{fofxy}
\end{eqnarray}
what shows that such an interpolation leads automatically to an
expansion in terms of factored functions of $x$ and $y$.
   Thus, Eq.~(\ref{phi}) is just the interpolation of a three-dimensional
function in its first two variables.
   In this latter case, however, the interpolation points $q_{\alpha}$ 
must be appropriate for every value of the third variable $r_{12}$.

   The fact that $r_{12}$ acts as a scaling factor (i.\ e.\ increasing
$r_{12}$ merely ``contracts'' $\phi$ as a function of $q_1$ and $q_2$,
without changing its shape) suggests a logarithmic mesh of points 
$q_{\alpha}$, in which 
$(q_{\alpha+1}-q_{\alpha}) = \lambda (q_{\alpha}-q_{\alpha-1})$,
with $\lambda > 1$.
   Such a logarithmic mesh is also suggested by the shape of 
$\phi(d_1,d_2)$ shown in Fig.~1 of ref.~[\onlinecite{Dion2004}].
   We have found that $N_{\alpha} \sim 20$ interpolation points 
$q_{\alpha}$ are sufficient for an accurate description of $\phi$
up to a cutoff $q_c$ at which we artificially ``saturate''
the original function $q_0(n,|\gn|)$ by redefining
\begin{equation}
  q_0^{sat}(n,|\gn|) = h[q_0(n,|\gn|),q_c]
\label{qsat}
\end{equation}
where $h(x,x_c)$ is a smooth function such that $h(x,x_c) \simeq x$ for 
$x<x_c$ and $h(x,x_c) \rightarrow x_c$ for $x \rightarrow \infty$:
\begin{equation}
  h(x,x_c) = x_c \left[ 1 - 
     \exp \left( -\sum_{m=1}^{m_c} \frac{(x/x_c)^m}{m} \right) \right]
\label{Sat}
\end{equation}
with $m_c \sim 12$ and $q_c \sim 5$ a.u.
   Higher $q_0$ values are obtained only for very large $n(\rr)$
(i.e. close to the nucleus, where $E_c^{nl}$ is negligible compared
to other terms in $E_{xc}$), and for large $|\gn|/n$
(in the electron density tails, where $E_c^{nl}$ is negligible
because of the factor $n(\rr)$ in the integrand of Eq.~(\ref{Enl})).
   In what follows, we will omit, but assume, superindex ``sat'' in
$q_0(n,|\gn|)$.

   A minor but significant difficulty is that $\phi(d_1,d_2)$ has
a logarithmic divergence when $d_1, d_2 \rightarrow 0$, what 
prevents its straightforward interpolation.
   Therefore, we interpolate and use instead a modified ``soft'' form
\begin{equation}
\phi_s(d_1,d_2) = \left\{ \begin{array}{ll} 
   \phi_0 + \phi_2 d^2 + \phi_4 d^4     & \mbox{if $d<d_s$} \\
   \phi(d_1,d_2)                        & \mbox{otherwise.}
  \end{array}
\right.
\label{PhiSoft}
\end{equation}
where $d = \sqrt{d_1^2+d_2^2}$. $\phi_0$ and $d_s$ are fixed parameters,
and $\phi_2, \phi_4$ are adjusted so that $\phi_s(d_1,d_2)$ and 
$\phi(d_1,d_2)$ match in value and slope at $d=d_s$ (for given $d_2/d_1$).
   This modification leads to a change in $E_c^{nl}$, which is 
corrected using a local density approximation:
\begin{equation}
\Delta E_c^{nl} = \int d^3\rr ~n(\rr) ~\Delta \epsilon_c^{nl}(\mathbf{r})
\label{DEnl}
\end{equation}
where
\begin{eqnarray}
\Delta \epsilon_c^{nl}(\rr) 
  &=& \frac{n(\rr)}{2} \int_0^\infty 4 \pi r'^2 dr'
   \left[ \phi(q,q,r') - \phi_s(q,q,r') \right]
   \nonumber \\
   &=& \frac{n(\rr)}{2 q^3} \int_0^{d_s} 4 \pi d^2 dd
    \left[ \phi(d,d) - \phi_s(d,d) \right]
\label{Depsnl}
\end{eqnarray}
with $q=q_0(n(\rr),\gn(\rr))$.
   The evaluation of $\Delta E_c^{nl}$ and its derivatives is performed, 
like that of the semilocal terms in Eq.~(\ref{Exc}), as in 
ref.~[\onlinecite{Balbas2001}].
   In what follows, we will assume, but omit for simplicity, 
the subindex $s$ in $\phi_s$.

   Substitution of (\ref{phi}) into (\ref{Enl}) leads to
\begin{eqnarray}
E_c^{nl} 
&=& \frac{1}{2} \sum_{\alpha \beta} \int\int d^3\rr_1 ~d^3\rr_2 
     ~\theta_{\alpha}(\rr_1) ~\theta_{\beta}(\rr_2) 
     ~\phi_{\alpha \beta}(r_{12}) \nonumber \\
&=& \frac{1}{2} \sum_{\alpha \beta} \int d^3\kk
    ~\theta_{\alpha}(\kk) ~\theta_{\beta}(\kk) ~\phi_{\alpha \beta}(k)
\label{Eoffk}
\end{eqnarray}
where $\theta_{\alpha}(\rr) = n(\rr) p_{\alpha}( q_0(n(\rr),\gn(\rr)) )$ 
and $\theta_{\alpha}(\kk)$ is its Fourier transform.
   Equally, $\phi_{\alpha \beta}(k)$ is the Fourier transform of
$\phi_{\alpha \beta}(r) \equiv \phi(q_{\alpha},q_{\beta},r)$.
   It can be calculated in spherical coordinates, and stored in a 
radial mesh of points $k$ for convenient interpolation.
   Thus, the heavier part of the calculation is the fast Fourier
transforms of the $N_\alpha$ functions $\theta_{\alpha}(\rr)$,
which still have a very moderate cost in a typical density
functional calculation.

   The evaluation of atomic forces requires the use of the 
Hellman-Feynman theorem, which holds only if the full energy functional 
is minimized selfconsistently.
   In turn, this requires the nonlocal part of the correlation 
potential, i.\ e.\ the functional derivative of Eq.~(\ref{Enl})
\cite{Thonhauser2007}.
   To handle the gradient dependence in $q_0(n,\gn)$ we use the 
same technique as in ref.~[\onlinecite{Balbas2001}]: approximating 
the spatial integrals by sums in a uniform grid of points, and the 
gradients by finite differences in the same grid.
   This makes $E_c^{nl}$ an ordinary function of the densities $n_i$
at fixed grid points $\rr_i$, allowing to perform conventional partial 
derivatives, rather than functional derivatives.
   Besides its conceptual simplicity, this method ensures a perfect 
consistency between the calculated potential and the energy:
\begin{equation}
 E_c^{nl} = \frac{1}{2} \Delta \Omega^2 \sum_{\alpha \beta} \sum_{ij} 
   \theta_{\alpha i} ~\theta_{\beta j} ~\phi_{\alpha \beta}(r_{ij})
\label{Eofni}
\end{equation}
where $\Delta \Omega$ is the volume per grid point and
$\theta_{\alpha i} \equiv n_i p_\alpha(q_0(n_i,\gn_i))$.
   Notice that $\phi_{\alpha \beta}(r_{ij})$ does not depend on
$n_i$, since the values $q_\alpha$ are fixed.
   A straightforward derivation then gives
\begin{eqnarray}
\lefteqn{ v_i^{nl} \equiv
     \frac{1}{\Delta \Omega } 
     \frac{\partial E_c^{nl}}{\partial n_i} } \nonumber \\
&& = \sum_{\alpha} \left( u_{\alpha i} 
     \frac{\partial \theta_{\alpha i}}{\partial n_i} + 
     \sum_j u_{\alpha j} 
          \frac{\partial \theta_{\alpha j}}{\partial \gn_j} 
          \frac{\partial \gn_j}{\partial n_i} \right)
\label{Vi}
\end{eqnarray}
where $\partial \gn_j / \partial n_i$ are fixed coefficients (determined
by the finite difference formula used for $\gn_j$) that depend only on
$\rr_{ij}$ and that are nonzero only for small $r_{ij}$.
   Also,
\begin{equation}
u_{\alpha i} = \Delta \Omega \sum_\beta \sum_j 
   \theta_{\beta j} ~\phi_{\alpha \beta}(r_{ij})
\label{epsilon}
\end{equation}
  is a convolution that can be obtained using fast Fourier 
transforms since (apart from $\pi$ and volume factors)
\begin{equation}
  \int d^3\rr_2 ~\theta_{\beta}(\rr_2) ~\phi_{\alpha \beta}(r_{12})
  = \int d^3\kk ~e^{i \kk \rr_1} ~\theta_{\beta}(\kk) ~\phi_{\alpha \beta}(k).
\label{epsofk}
\end{equation}
   Thus, a selfconsistency step requires $N_{\alpha}$ direct transforms
to find $\theta_{\alpha}(\kk)$ and $N_{\alpha}$ inverse transforms to obtain
$u_{\alpha}(\rr)$.
   The calculation of the atomic forces does not require any additional
effort, since the nonlocal contribution $v_i^{nl}$ is simply added to 
the semilocal terms~\cite{Balbas2001} in $v_i^{xc}$ and to the rest of 
the effective potential.
   Notice that the implementation is independent of the basis set,
accepting $n(\rr_i)$ in a uniform real space grid $\rr_i$ and returning 
$E_{xc}$ and $v_{xc}(\rr_i)$ in the same grid.
   It has been checked that it reproduces accurately the results
obtained by direct evaluation of Eq.~(\ref{Enl}) (and, eventually, 
its functional derivative~\cite{Thonhauser2007}) for a variety
of systems \cite{Dion2004,Thonhauser2006}.
 
   We have applied the above method to study the interaction
between the concentric layers of double-wall carbon nanotubes (DWNT).
   Such interactions are crucial for different nanodevices proposed 
recently~\cite{Cumings-Zettl2000,Barreiro2008} and they have been
studied with semiempirical potentials~\cite{Saito2001} and with a local 
DFT functional 
\cite{Charlier-Michenaud1993,Bichoutskaia2005,Bichoutskaia2006}.
   We have used the SIESTA code~\cite{Ordejon-Artacho-Soler1996,Soler2002} 
with an optimized~\cite{Anglada2002} triple-$\zeta$+polaratization 
basis set of pseudoatomic orbitals, correcting for basis set superposition 
errors (BSSE).
   The integration grids in real and reciprocal space had cutoffs
of 300 Ry and 20 \AA, respectively.
   The cutoff parameter for \textit{k}-point sampling 
\cite{Moreno-Soler1992} was 20~\AA, ensuring at least 34, 20 and 14  
\textit{k}-points for the armchair, zigzag and chiral DWNTs studied,
respectively.
   The atomic forces were relaxed to less than 20 meV/\AA.

\begin{figure}[htpb]
\includegraphics[width=0.9\columnwidth,clip]{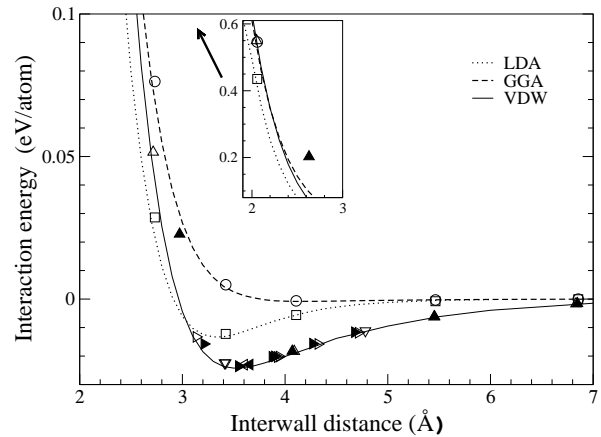}
\caption{ 
   Interaction and formation energies between different double wall carbon 
nanotubes, as a function of interwall separation, using 
LDA~\cite{Perdew-Zunger1981} (squares), GGA~\cite{Perdew-Burke-Ernzerhof1996} 
(circles), and van der Waals~\cite{Dion2004} (triangles) functionals.
   Interaction energies (empty symbols) are between two individually
   relaxed rigid tubes.
   Formation energies (filled symbols) include also the geometry relaxation 
induced by the interaction, that modifies their interwall separation.
   The tube geometries are (5,5)@(n,n) ($\bigcirc,\Box,\triangle)$),
(m,m)@(n,n) m$>$5 ($\bigtriangledown$), (m,0)@(n,0) ($\rhd$), and
(8,2)@(16,4) ($\lhd$).
   For comparison, we also show the interaction energies for two flat 
graphene layers (lines).
   All energies are divided by the total number of atoms in both tubes.
}
\label{interaction_energy}
\end{figure}

   Figure \ref{interaction_energy} shows the calculated interaction 
energy between two rigid SWNTs, relaxed independently, as a function 
of their interwall separation (difference of radii). 
   It also shows the DWNT formation energies, defined as the difference
between the total energy of the relaxed DWNT and that of the two SWNTs.
   The calculated tubes (m,m)@(n,n) (armchair), 
(m,0)@(n,0) (zigzag) and (8,2)@(16,4) (chiral) were chosen for 
their conmensurability in the longitudinal direction, as well 
as for comparison with prior calculations.
   It can be seen that the interaction energy depends neglegibly
on chirality and curvature, being very well represented by 
the interaction between two flat graphene layers.
   On the other hand, the formation energy, that includes the
relaxation of the radii induced by the interaction, shows a steeper 
repulsion than between flat graphene layers. 
   In agreement with previous results for graphene and graphite, we
find that the LDA works reasonably well.
   For sufficiently long tubes, in which the border effects can be
neglected, the calculated vdW interaction energy gives a telescopic
contraction force~\cite{Cumings-Zettl2000} $F = 0.91 $N/m $\times ~d$, 
where $d$ is the mean of the inner and outer tube diameters.

   Next, the two concentric tubes of the DWNT were moved rigidly, relative to 
each other, in order to construct rotation-translation energy maps.
   To generate these maps, we first project the inner tube 
coordinates onto the outer tube surface, i.\ e.\ we multiply its
$x$ and $y$ coordinates (the tube axis being $z$) by the ratio 
$R_{out}/R_{in}$ between the two radii.
   We then unroll the coordinates of both tubes onto a flat surface, 
repeating them periodically also in the $x$ axis.
   This gives two flat periodic lattices (conmensurate in the
cases considerd) with reciprocal unit cell vectors $\aa_i$
and $\bb_i, i=1,2$.
   The energy maps can then be represented, as a function of the
position $\xx$ on this surface, relative to the minimum, by an 
expansion of the form
\begin{equation}
  U(\xx) = U_0 - \frac{1}{4} \sum_{\GG \ne 0} U_{\GG} 
                 \cos(\GG \cdot \xx)
\label{Uofx}
\end{equation}
where $\GG$ are the superlattice wavevectors, common to the reciprocal 
lattices $\aa$ and $\bb$, and $U_{\GG}$ are the barrier heights for
motion along $\GG$.
   We have found that limiting this expansion to the first two 
wavevector stars, $\pm \GG_1$ and $\pm \GG_2$ (which, in the cases 
studied, are parallel and orthogonal to the axial direction), 
gives a good approximation to the cases studied, with the parameters 
given in Table~\ref{barriers}.
\begin{table}
\begin{tabular}{lccc}
\hline\hline
DWNT             & (5,5)@(10,10) & (9,0)@(18,0) & (8,2)@(16,4) \\
\hline
$\Delta x_z$     &     1.24      &     2.15     &     0.47     \\
$\Delta x_\phi$  &     2.15      &     1.24     &     0.81     \\
$U_z^{LDA}$      &     0.07      &     1.38     &     0.00     \\   
$U_\phi^{LDA}$   &     0.48      &     0.16     &     0.00     \\
$U_z^{vdW}$      &     0.04      &     1.22     &     0.00     \\
$U_\phi^{vdW}$   &     0.43      &     0.06     &     0.00     \\
\hline\hline
\end{tabular}
\caption{
   Periodicities ($\Delta x_i = 2 \pi / G_i$, in \AA) and energy barriers 
$U_i$ (in meV per outer tube atom) for translation ($i=z$)
and rotation ($i=\phi$) of the outer tube, relative to the inner tube, 
in double wall carbon nanotubes.
  $\Delta x_\phi$ lengths are along the outer tube circunference.
  For the (8,2)@(16,4) tube we found that all the LDA and vdW barriers
are smaller than out computational accuracy of $\sim 0.01$ meV/atom.
}
\label{barriers}
\end{table}

   Overall, the relative values of these barrier heights are in
qualitative agreement with previous calculations, i.\ e.\ larger
barrier distances lead to larger barrier heights.
   Quantitatively, however, those calculations vary by an order
of magnitude depending on the models 
used~\cite{Saito2001,Bichoutskaia2005}.
   Our calculated LDA barriers are similar to those of 
refs.~\cite{Charlier-Michenaud1993,Bichoutskaia2005,Bichoutskaia2006}.
   The small discrepancies with ref.~\cite{Bichoutskaia2006} may be due
to the different basis sets and to our finer $k$-point sampling.
   Again, we find that the LDA does a rather good job for these systems,
compared to the more complex vdW functional.  
   Nevertheless, we find that LDA systematically underestimates the 
interaction energies and that it overestimates the barrier heights,
relatively to the vdW results.
 
   In conclusion, we have described an efficient algorithm to include
van de Waals interactions through the selfconsistent treatment of
a nonlocal {\it ab initio} functional proposed recently.~\cite{Dion2004}
   Typical overheads in the total computation time, using the
SIESTA code, over that required by LDA or GGA, 
are a factor $\sim 5$ in a two-atom system and a $\sim 10 \%$ 
increase for $\sim 150$ atoms.
   Using this implementation, we have calculated the interaction  
energies, as well as the barriers for relative displacement,
between concentric tubes in several armchair, zigzag, and chiral 
DWNTs.

   We would like to thank E.\ Anglada, E.\ Artacho, and D.\ C.\ Langreth 
for many discussions and for their help in generating basis sets and 
testing our implementation.
   This work has been founded by grant FIS2006-12117 from the
Spanish Ministery of Science.

\bibliographystyle{apsrev}
\bibliography{dft,siesta,vdw}

\end{document}